\documentclass[a4paper,11pt]{article}
\usepackage{graphicx,epsfig,picins}
\usepackage{fancyhdr,fancybox}
\usepackage{indentfirst}
\usepackage{titlesec}
\usepackage{verbatim}
\usepackage[sort&compress, numbers]{natbib}
\usepackage{array,dcolumn,tabularx}
\usepackage{setspace}
\usepackage{geometry}
\usepackage{extarrows,chemarrow,xypic} 
\usepackage[small]{caption2}
\usepackage{lineno}
\usepackage{microtype}
\DisableLigatures[f]{encoding = *, family = * }

\usepackage{times}     

\usepackage{latexsym}
\usepackage{amsmath}                 
\usepackage{amssymb}                 
\usepackage{amsbsy}
\usepackage{amsthm}
\usepackage{amsfonts}
\usepackage{mathrsfs}                
\usepackage{bm}                      
\usepackage{arydshln}
\usepackage{relsize}                 
\usepackage{hyperref}  

\newcommand{\paperfont}{\fontsize{12pt}{1.3\baselineskip}\selectfont}
\begin{document}

\theoremstyle{definition}
\makeatletter
\thm@headfont{\bf}
\makeatother
\newtheorem{definition}{Definition}
\newtheorem{example}{Example}
\newtheorem{theorem}{Theorem}
\newtheorem{lemma}{Lemma}
\newtheorem{corollary}{Corollary}
\newtheorem{remark}{Remark}
\newtheorem{proposition}{Proposition}

\lhead{}
\rhead{}
\lfoot{}
\rfoot{}

\renewcommand{\refname}{References}
\renewcommand{\figurename}{Fig.}
\renewcommand{\tablename}{Table}
\renewcommand{\proofname}{Proof}

\newcommand{\diag}{\mathrm{diag}}
\newcommand{\tr}{\mathrm{tr}}
\newcommand{\dnum}{\mathrm{d}}
\newcommand{\Enum}{\mathbb{E}}
\newcommand{\Pnum}{\mathbb{P}}
\newcommand{\Rnum}{\mathbb{R}}
\newcommand{\Cnum}{\mathbb{C}}
\newcommand{\Znum}{\mathbb{Z}}
\newcommand{\Nnum}{\mathbb{N}}
\newcommand{\abs}[1]{\left\vert#1\right\vert}
\newcommand{\set}[1]{\left\{#1\right\}}
\newcommand{\norm}[1]{\left\Vert#1\right\Vert}
\newcommand{\Q}{\boldsymbol{Q}}
\newcommand{\W}{\boldsymbol{W}}
\newcommand{\I}{\boldsymbol{I}}
\newcommand{\M}{\boldsymbol{M}}
\newcommand{\p}{\boldsymbol{p}}
\newcommand{\pai}{\boldsymbol{\pi}}


\title{\textbf{Simplification of Markov chains with infinite state space and the mathematical theory of random gene expression bursts}}
\author{Chen Jia$^{1}$ \\
\footnotesize $^1$Department of Mathematical Sciences, University of Texas at Dallas, Richardson, TX 75080, U.S.A.\\
\footnotesize Email: jiac@utdallas.edu}
\date{}                              
\maketitle                           
\thispagestyle{empty}                

\paperfont

\begin{abstract}
Here we develop an effective approach to simplify two-time-scale Markov chains with infinite state spaces by removal of states with fast leaving rates, which improves the simplification method of finite Markov chains. We introduce the concept of fast transition paths and show that the effective transitions of the reduced chain can be represented as the superposition of the direct transitions and the indirect transitions via all the fast transition paths. Furthermore, we apply our simplification approach to the standard Markov model of single-cell stochastic gene expression and provide a mathematical theory of random gene expression bursts. We give the precise mathematical conditions for the bursting kinetics of both mRNAs and proteins. It turns out that random bursts exactly correspond to the fast transition paths of the Markov model. This helps us gain a better understanding of the physics behind the bursting kinetics as an emergent behavior from the fundamental multi-scale biochemical reaction kinetics of stochastic gene expression.
\\

\end{abstract}

\section{Introduction}
Markov chains with finite or infinite state spaces play a crucial role in the stochastic modeling of biochemical systems and networks. Finite Markov chains are widely applied to model single-molecule enzyme kinetics \cite{qian2002single}, allostery of receptors or ion channels \cite{jia2014allosteric}, phenotypic switching of cell populations \cite{gupta2011stochastic}, etc. Infinite Markov chains are equally important because they act as the mathematical foundation of mesoscopic stochastic biochemical reaction kinetics \cite{kurtz1972relationship, anderson2015stochastic}. In fact, all cellular events directly or indirectly depend on stochastic collisions between various biochemical molecules. Assume that a biochemical reaction network involves $M$ biochemical species whose copy numbers are denoted by $n_1,\cdots,n_M$. Then the network can be model by a huge and infinite Markov chain on the $M$-dimensional lattice: $\{(n_1,\cdots,n_M):n_1,\cdots,n_M\geq 0\}$ whose evolution is governed by the famous chemical master equation, which is first introduced in the work of Leontovich \cite{leontovich1935basic} and Delbr\"{u}ck \cite{delbruck1940statistical}. This explains why infinite Markov chains are so important in chemistry and biology.

In applications, the Markov model of a biochemical system often possesses two or multiple different time scales. This raises a natural question of whether we can simplify the Markov model to a simpler one with very little loss of the dynamic information. The simplification of Markov chains is important in many ways. From the mathematical perspective, it allows us to perform an in-depth theoretical analysis of the model. From the statistical perspective, it enables us to obtain a robust estimation of the model parameters and improves the statistical significance of data fitting. From the physical perspective, it helps us gain a clearer understanding of the dynamic features and emergent behavior of the system.

The simplification of two-time-scale finite Markov chains have been extensively studied. In general, there are two types of simplification strategies: removal of transient states, also called decimation, and aggregation of recurrent states, also called averaging \cite{jia2016model, bo2016multiple}. In this paper, we focus on the previous strategy. In fact, the decimation strategy for finite Markov chains dated back to the study on stiff Markov chains by engineers \cite{bobbio1986aggregation, reibman1989analysis, bobbio1990computing, malhotra1994stiffness} and have been recently developed into a complete theory \cite{pigolotti2008coarse, yin2012continuous, ullah2012simplification, jia2016simplification}. It has been shown that a reversible Markov chain is equivalent to an resistor-capacitor circuit and the decimation strategy corresponds to the $Y-\Delta$ transformation in the circuit theory \cite{ullah2012simplification}. The decimation strategy is also generalized to irreversible Markov chains by removal of the fast states one by one \cite{pigolotti2008coarse, jia2016simplification} and the relationship between model simplification and irreversibility is also discussed in detail \cite{puglisi2010entropy, jia2016model}. However, the traditional decimation strategy becomes less effective when the number of fast states is infinite.

In this paper, we develop an effective simplification method for infinite Markov chains, which improves the traditional decimation strategy. This is the first main subject of this paper. In fact, an infinite Markov chain can be simplified to a reduced one by removal of the fast states all together. By introducing the concept of fast transition paths which has important biophysical implications, we show that the effective transition rates between any pair of states are the sum of the direct transition rates and the contribution of indirect transitions via all the fast transition paths. Similarly, the effective initial probability of any state is the superposition of the original initial probability and the contribution of all the fast transition paths.

In recent years, the stochastic modeling of gene expression within single cells has become one of the most quantitative aspect of molecular biology \cite{peccoud1995markovian, paulsson2000stochastic, kepler2001stochasticity, paulsson2005models, hornos2005self, friedman2006linking, raj2006stochastic, shahrezaei2008analytical, kumar2014exact, jia2014modeling, lin2016gene, jia2017stochastic}. To validate the effectiveness of our simplification approach, we apply it to the standard Markov model of single-cell stochastic gene expression involving gene switching, transcription, and translation \cite{paulsson2005models, shahrezaei2008analytical}. Recent single-cell experiments \cite{cai2006stochastic, cai2008frequency, suter2011mammalian} have shown that the synthesis of many mRNAs and proteins in a living cell may occur in random bursts | short periods of high expression intensity followed by long periods of low expression intensity \cite{paulsson2000stochastic}. Although there have been some intuitive understanding on how bursts occur, it is interesting to bring the bursting kinetics into a rigorous mathematical framework.

In this work, we present a mathematical theory of random gene expression bursts. This is the second main subject of this paper. We show that both the transcription and translation bursts naturally emerge from the fundamental multi-scale biochemical reaction kinetics of stochastic gene expression. It turns out that random bursts precisely correspond to fast transition paths of the standard Markov model. Furthermore, we also give the precise mathematical conditions for the bursting kinetics of both mRNAs and proteins, which are in full agreement with recent bulk and single-cell experiments.

\section{Simplification of infinite Markov chains}

\subsection{Model and previous results}
We consider a inhomogeneous continuous-time Markov chain with infinite state space $S = \{1,2,\cdots\}$, where the states are labeled by positive integers. Let $\Q(t) = (q_{ij}(t))$ denote the generator matrix of the chain, where $q_{ij}(t)$ with $i\neq j$ denotes the transition rate from state $i$ to state $j$ at time $t$ and $q_{ii}(t) = -\sum_{j\neq i}q_{ij}(t)$. Following standard notations, let
\begin{equation*}
q_i(t) = -q_{ii}(t) = \sum_{j\neq i}q_{ij}(t)
\end{equation*}
denote the rate at which the chain leaves state $i$ at time $t$. Let $\pai = (\pi_i)$ denote the initial distribution of the chain and let $\p(t) = (p_i(t))$ denote the probability distribution of the chain at time $t$. Then the evolution of the probability distribution is governed by the master equation
\begin{equation}\label{master}\left\{
\begin{split}
\dot{\p}(t) &= \p(t)\Q(t), \\
\p(0) &= \pai.
\end{split}\right.
\end{equation}

We further assume that the generator matrix $\Q(t)$ depends on a single parameter $\lambda$. When $\lambda\gg 1$, some states have relatively fast leaving rates compared to other states and thus the chain will possess two separate times scales. We next introduce an important definition. If
\begin{equation*}
\lim_{\lambda\rightarrow\infty}q_i(t) = \infty,\;\;\;\forall t\geq 0,
\end{equation*}
then state $i$ has a relatively fast leaving rate compared to other states and is referred to as a \emph{fast state}. In contrast, if
\begin{equation*}
\lim_{\lambda\rightarrow\infty}q_i(t) < \infty,\;\;\;\forall t\geq 0,
\end{equation*}
then $i$ is called a \emph{slow state}. Intuitively, if $i$ is a fast state, then the time that the chain stays in this state will be very short and thus we may except the chain to be simplified by removal of the fast states. Let $A$ denote the set of all the slow states and let $B$ denote that of all the fast states\footnote{Here we assume that the state space $S$ is the union of $A$ and $B$. In other words, we assume that any state is either a fast one or a slow one.}. By relabelling the states, the initial distribution can be represented as $\pai = (\pai_A,\pai_B)$, the probability distribution can be represented as $\p(t) = (\p_A(t),\p_B(t))$, and the generator matrix can be represented as
\begin{equation*}
\Q(t) = \begin{pmatrix} \Q_{AA}(t) & \Q_{AB}(t) \\ \Q_{BA}(t) & \Q_{BB}(t) \end{pmatrix}.
\end{equation*}

Most of the previous papers on model simplification assumed that \\
(a) the Markov chain is finite;\\
(b) the Markov chain is homogeneous, that is, the generator matrix $\Q$ does not depend on time $t$;\\
(c) the generator matrix has the following singularly perturbed form:
\begin{equation*}
\Q = \lambda\tilde\Q+\hat\Q.
\end{equation*}
where $\tilde\Q$ and $\hat\Q$ are two generator matrices governing the fast and slow processes, respectively. However, these assumptions are often violated for mesoscopic biochemical reaction kinetics (see Sec. \ref{discussion}). In this paper, we consider a more general case by removing these three assumptions.

For finite homogenous Markov chains, previous studies \cite{pigolotti2008coarse, jia2016simplification} have shown that the original chain can be simplified to a reduced one by removal of the fast states. Specifically, the state space of the reduced chain is the slow state space $A$, the generator matrix $\tilde{\Q} = (\tilde{q}_{ij})$ of the reduced chain is given by
\begin{equation*}
\tilde{\Q} = \Q_{AA}-\Q_{AB}\Q_{BB}^{-1}\Q_{BA},
\end{equation*}
and the initial distribution $\tilde\pai = (\tilde\pi_i)$ of the reduced chain is given by
\begin{equation*}
\tilde{\pai} = \pai_A-\pai_{B}\Q_{BB}^{-1}\Q_{BA}.
\end{equation*}
If the chain has an infinite number of fast states, however, the infinite-dimensional matrix $Q_{BB}^{-1}$ is not well defined and is hard to compute. This suggests the necessity to generalize the above formulas to the case of infinite state space.

\subsection{Generator matrix of the reduced chain}
Assume that the Markov chain jumps from state $i$ to another state at time $t$. When $\lambda\gg 1$, the transition probability from state $i$ to state $j$ at time $t$ is given by
\begin{equation*}
w_{ij}(t) = \lim_{\lambda\rightarrow\infty}\frac{q_{ij}(t)}{q_i(t)},\;\;\;\forall j\neq i.
\end{equation*}
Let $\W(t) = (w_{ij}(t))$ denote the transition probability matrix, also called the jump matrix, of the chain at time $t$, where $w_{ii}(t) = 0$ for any state $i$. We also represent the transition probability matrix as
\begin{equation*}
\W(t) = \begin{pmatrix} \W_{AA}(t) & \W_{AB}(t) \\ \W_{BA}(t) & \W_{BB}(t) \end{pmatrix}.
\end{equation*}
For convenience, let $\M_B(t) = \diag(q_i(t))$ denote the diagonal matrix whose diagonal entries are the leaving rates $q_i(t)$ with $i$ ranging over all the fast states. It is easy to see that for any pair of states $i$ and $j$,
\begin{equation*}
q_{ij}(t) = q_i(t)(w_{ij}(t)-\delta_{ij}),
\end{equation*}
where $\delta_{ij}$ is Kronecker's delta function. This implies that
\begin{equation*}
\begin{split}
\Q_{BA}(t) &= \M_B(t)\W_{BA}(t),\\
\Q_{BB}(t) &= -\M_B(t)(\I-\W_{BB}(t)).
\end{split}
\end{equation*}
Thus Eq. \eqref{master} can be rewritten as
\begin{equation}\label{master1}
\begin{split}
\dot{\p}_A(t) &= \p_A(t)\Q_{AA}(t)+\p_B(t)\M_B(t)\W_{BA}(t), \\
\dot{\p}_B(t) &= \p_A(t)\Q_{AB}(t)-\p_B(t)\M_B(t)(\I-\W_{BB}(t)).
\end{split}
\end{equation}
In the above equation, $\p_A(t)$ and $\p_B(t)$ are the slow and fast variables, respectively. If we focus on the fast variable, then the slow variable can be frozen. From Eq. \eqref{master1}, the quasi-steady state of the fast variable is given by
\begin{equation}\label{averaging}
\p_B(t)_{qss} = \p_A(t)\Q_{AB}(t)(\I-\W_{BB}(t))^{-1}\M_B(t)^{-1}.
\end{equation}
On the other hand, if we focus on the slow variable, we can think that the fast variable has reached the quasi-steady state. From Eqs. \eqref{master1} and \eqref{averaging}, the dynamics of the slow variable is reduced to
\begin{equation*}
\begin{split}
\dot{\p}_A(t) &= \p_A(t)\Q_{AA}(t)+\p_B(t)_{qss}\M_B(t)\W_{BA}(t) \\
&= \p_A(t)[\Q_{AA}(t)+\Q_{AB}(t)(\I-\W_{BB}(t))^{-1}\W_{BA}(t)].
\end{split}
\end{equation*}
This indicates that the original chain can be simplified to a reduced one by removal of the fast states and the generator matrix of the reduced chain is given by
\begin{equation}\label{effective1}
\tilde{\Q}(t) = \Q_{AA}(t)+\Q_{AB}(t)(\I-\W_{BB}(t))^{-1}\W_{BA}(t).
\end{equation}

\begin{figure}[!htb]
\begin{center}
\centerline{\includegraphics[width=0.4\textwidth]{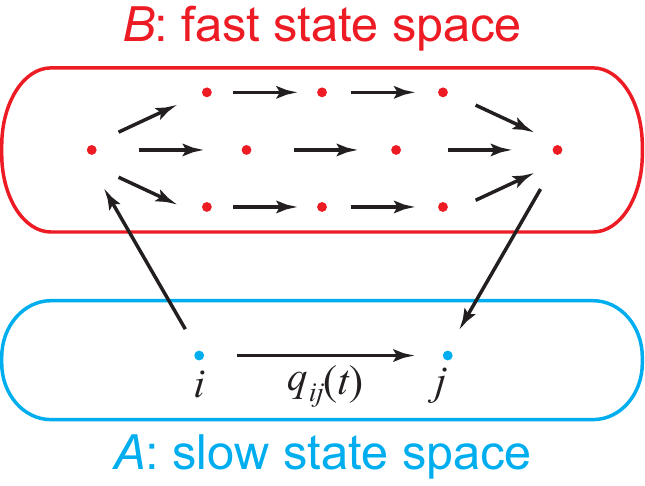}}
\caption{\textbf{Schematic diagram of the simplification method.} The effective transition rate from state $i$ to state $j$ is the sum of the direct transition rate and the contribution of indirect transitions via all the fast transition paths.}\label{decimation}
\end{center}
\end{figure}

Even if the chain has an infinite number of states, the matrix $(\I-\W_{BB}(t))^{-1}$ is still well defined as the following infinite series:
\begin{equation*}
(\I-\W_{BB}(t))^{-1} = \sum_{n=0}^\infty\W_{BB}(t)^n.
\end{equation*}
In applications, we always hope to calculate this infinite series. To this end, we introduce another key concept. Let $i_1,i_2,\cdots,i_n$ be a sequence of states. We say that $c: i_1\rightarrow i_2\rightarrow\cdots\rightarrow i_n$ is a \emph{fast transition path} from state $i$ to state $j$ if $i_1 = i$, $i_n = j$, and the intermediate states $i_2,\cdots,i_{n-1}$ are all fast states. Moreover, the probability weight of the fast transition path $c$ at time $t$ is defined as
\begin{equation*}
w_c(t) = w_{i_1i_2}(t)w_{i_2i_3}(t)\cdots w_{i_{n-1}i_n}(t).
\end{equation*}
It is easy to see that in order that the fast transition path $c$ has a positive probability weight, all the intermediate transitions along this path must satisfy
\begin{equation}\label{observation}
\lim_{\lambda\rightarrow\infty}q_{i_2i_3}(t) = \cdots = \lim_{\lambda\rightarrow\infty}q_{i_{n-1}i_n}(t) = \infty.
\end{equation}
It is always convenient to find all the fast transition paths with positive probability weights by using the above criterion. From Eq. \eqref{effective1}, the effective transition rate from state $i$ to state $j$ at time $t$ can be written as
\begin{equation}\label{superposition}
\begin{split}
\tilde{q}_{ij}(t) &= q_{ij}(t)+\sum_{k,l\in B}q_{ik}(t)[\sum_{n=0}^\infty W_{BB}(t)^n]_{kl}w_{lj}(t)\\
&= q_{ij}(t)+q_i(t)\sum_cw_c(t),
\end{split}
\end{equation}
where $c$ ranges over all the fast transition paths from state $i$ to state $j$. This formula indicates that the effective transition rate $\tilde{q}_{ij}(t)$ from state $i$ to state $j$ is the sum of two parts: the first part is the direct transition rate $q_{ij}(t)$ and the second part is the contribution of indirect transitions via fast transition paths, as illustrated in Fig. \ref{decimation}.

We stress here that if a fast transition path $i\rightarrow k\rightarrow\cdots\rightarrow j$ starts from a slow state, then the first transition $i\rightarrow k$ is a slow process and the remaining transitions are all fast processes which are essentially instantaneous. Therefore, the direct and indirect transitions have the same time scale. Specifically, from Eq. \eqref{superposition}, the direct and indirect transition rates from state $i$ to state $j$ are given by
\begin{equation*}
\begin{split}
q_{ij}^{\textrm{direct}}(t) = q_i(t)w_{ij}(t),\;\;\;q_{ij}^{\textrm{indirect}}(t) = q_i(t)\sum_cw_c(t).
\end{split}
\end{equation*}
In addition, we have the following inequality:
\begin{equation*}
\begin{split}
w_{ij}(t)+\sum_cw_c(t)\leq 1.
\end{split}
\end{equation*}
This can be understood as follows. Since $\W(t)$ is a transition probability matrix, it corresponds to a discrete-time Markov chain. The left-hand side of the above inequality is less than or equal to the probability that this discrete-time Markov chain jumps from state $i$ to state $j$ within finite number of transitions, which is less than or equal to 1. Thus the two terms $w_{ij}(t)$ and $\sum_cw_c(t)$ are both quantities between 0 and 1. This again shows that the direct and indirect transitions have the same time scale.

\subsection{Initial distribution of the reduced chain}
If the original chain starts from a slow state, then the reduced chain should of course start from the same state. If the original chain starts from a fast state, however, the reduced chain cannot start from the same state because the fast states are not included in the state space of the reduced chain. This raises a natural question of how the initial distribution of the effective chain should be chosen.

Let $\tau_A$ denote the first passage time of the slow state space $A$, which is the time needed for the chain to reach $A$ for the first time. Intuitively, if the original chain starts from a fast state, it will jump to a slow state within a very short period. This suggests that the initial distribution of the reduced chain should agree with the probability distribution of the original chain at time $\tau_A$, also called the first passage distribution of $A$. For homogeneous Markov chains for which the generator matrix $\Q$ is independent of time $t$, it is a classical result \cite{chung1967markov} that the first passage distribution of $A$ is given by
\begin{equation*}
\tilde\pai = \pai_A-\pai_B\Q_{BB}^{-1}\Q_{BA} = \pai_A+\pai_B(\I-\W_{BB})^{-1}\W_{BA}.
\end{equation*}
For inhomogeneous Markov chains, since the original chain will concentrate on the slow state space $A$ within a very short period, the initial distribution of the reduced chain should be chosen as
\begin{equation}\label{initial1}
\tilde\pai = \pai_A+\pai_B(\I-\W_{BB}(0))^{-1}\W_{BA}(0).
\end{equation}
Similarly, the effective initial probability of state $i$ can be represented by fast transition paths as
\begin{equation*}
\begin{split}
\tilde\pi_i &= \pi_i+\sum_{j,k\in B}\pi_j[\sum_{n=0}^\infty\W_{BB}(0)^n]_{jk}w_{ki}(0)\\
&= \pi_i+\sum_{j\in B}\pi_j\sum_cw_c(0),
\end{split}
\end{equation*}
where $c$ ranges over all the fast transition paths from state $j$ to state $i$. This formula indicates that the effective initial probability $\tilde\pi_i$ of state $i$ is the sum of two parts: the first part is the original initial probability $\pi_i$ and the second part is the contribution of fast transition paths.

If the original chain starts from a slow state, the probability distributions of the two chains will agree with each other over the whole time axis. However, if the original chain starts from a fast state, it will concentrate on the slow state space $A$ within a very short period. Once the original chain concentrates on $A$, the probability distributions of the two chains begin to agree with each other. In other words, although our simplification approach may cause large errors in the short-time regime, the probability distributions of the two chains will coincide with each other in the long-time regime.

\section{Mathematical theory of random gene expression bursts}

\subsection{Standard three-stage model of stochastic gene expression}
\begin{figure}[!htb]
\begin{center}
\centerline{\includegraphics[width=1.0\textwidth]{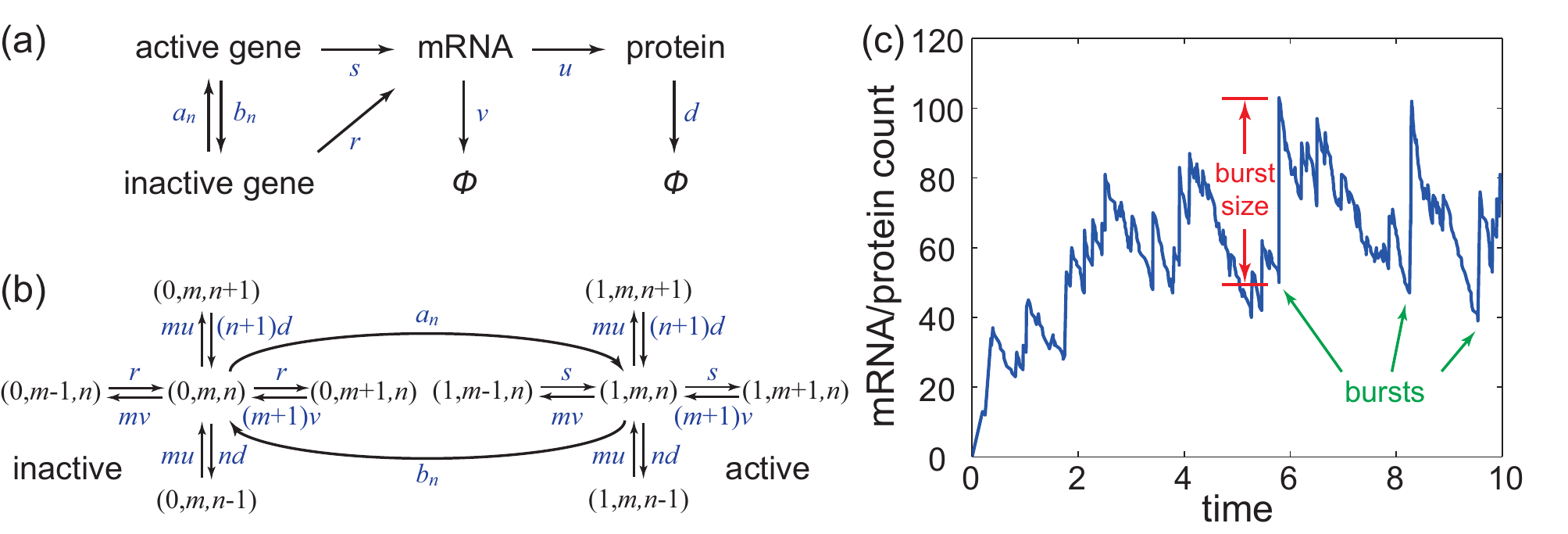}}
\caption{\textbf{Stochastic gene expression with the bursting kinetics.} (a) The standard model of single-cell stochastic gene expression with three stages: gene switching, transcription, and translation. (b) The transition diagram of the standard Markov model. (c) The schematic diagram of transcription or translation bursts.}\label{model}
\end{center}
\end{figure}

In recent years, the stochastic modeling of single-cell gene expression has become one of the most quantitative aspect of biophysics \cite{peccoud1995markovian, paulsson2000stochastic, kepler2001stochasticity, paulsson2005models, hornos2005self, friedman2006linking, raj2006stochastic, shahrezaei2008analytical, kumar2014exact, lin2016gene}. Based on the central dogma of molecular biology, the kinetics of stochastic gene expression in a single cell can described by a standard model with three stages involving gene switching between an active and an inactive states, transcription, and translation, as illustrated in Fig. \ref{model}(a) \cite{paulsson2005models, shahrezaei2008analytical}. The chemical state of the gene of interest can be described by three variables $(i,m,n)$: the gene activity $i$, the mRNA copy number $m$, and the protein copy number $n$. Here $i = 1$ and $i = 0$ correspond to the active and inactive states of the gene, respectively. Let $p^i_{mn}(t)$ denote the probability of having $m$ mRNAs and $n$ proteins at time $t$ when the gene is in state $i$. Then the dynamics of the standard model can be described by the continuous-time Markov chain with infinite state space illustrated in Fig. \ref{model}(b), whose evolution is governed by the chemical master equation
\begin{equation*}\left\{
\begin{split}
\dot p^1_{m,n} =&\; a_np^0_{m,n}+sp^1_{m-1,n}+mup^1_{m,n-1}\\
&\; +(m+1)vp^1_{m+1,n}+(n+1)dp^1_{m,n+1}\\
&\; -(b_n+s+mv+mu+nd)p^1_{m,n}, \\
\dot p^0_{m,n} =&\; b_np^1_{m,n}+rp^0_{m-1,n}+mup^0_{m,n-1}\\
&\; +(m+1)vp^0_{m+1,n}+(n+1)dp^0_{m,n+1}\\
&\; -(a_n+r+mv+mu+nd)p^0_{m,n}.
\end{split}\right.
\end{equation*}
Here $a_n$ and $b_n$ are the gene switching rates; $s$ and $r$ are the transcription rates when the gene is active and inactive, respectively; $u$ is the translation rate; $v$ and $d$ are the degradation rates of the mRNA and protein, respectively. In living cells, the products of many genes also regulate their own expression to form an autoregulatory gene network. This suggests that the gene switching rates $a_n$ and $b_n$ generally depend on the protein copy number $n$.

We stress here that, in theory, our simplification method can be applied to the case when all the reaction rates are time-dependent. However, for simplicity, we only focus on the case when all the reaction rates are time-independent, as assumed in previous papers.

\subsection{Emergent transcription bursts}
In single-cell experiments \cite{cai2006stochastic, cai2008frequency, suter2011mammalian}, it was frequently observed that many mRNAs or proteins could yield short periods of high expression intensity followed by long periods of low expression intensity. This phenomenon is called random bursts, whose schematic diagram is depicted in Fig. \ref{model}(c). The bursting kinetics is a phenomenological concept with two separate time scales: fast accumulation of mRNAs or proteins and slow decay. However, it was pointed out in \cite{paulsson2005models} that bursts should not be expected for most genes. This raises a natural question of what are the mathematical conditions for the bursting kinetics of mRNAs and proteins. In fact, the biochemical reactions involved in gene expression have multiple different time scales,
spanning many orders of magnitude \cite{moran2012sizing}. Here we shall apply the above simplification method to present a mathematical theory for random gene expression bursts.

\begin{figure}[!htb]
\begin{center}
\centerline{\includegraphics[width=0.8\textwidth]{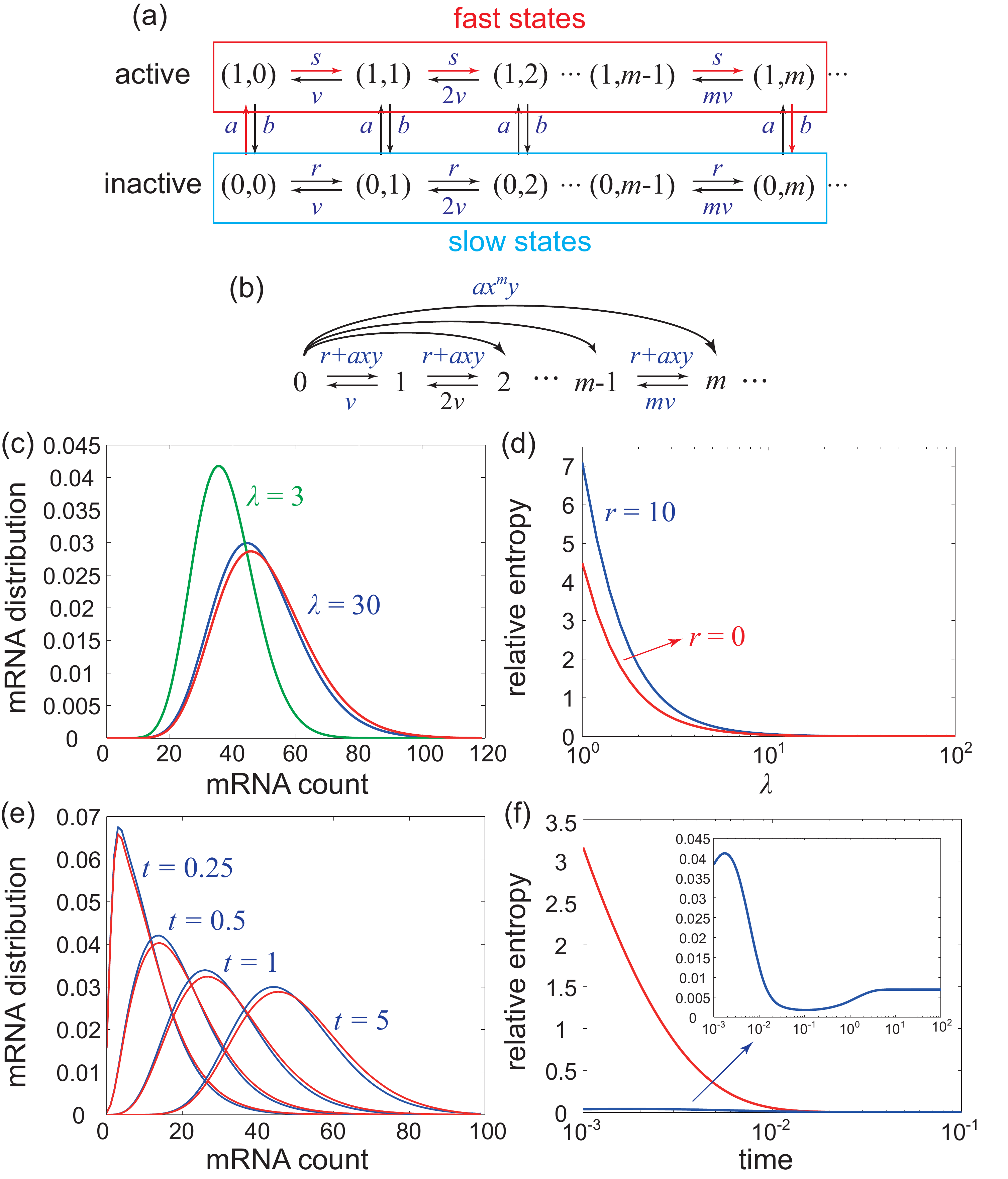}}
\caption{\textbf{Transcription bursts as an emergent behavior of the fundamental biochemical reaction kinetics.}
(a) The standard model with gene switching and transcription.
(b) The transition diagram of the reduced model when $b\gg a$ and $s/b$ is finite.
(c) Simulations of the steady-state mRNA distributions for the reduced model (red) and the original model when $\lambda = 3$ (green) and $\lambda = 30$ (blue).
(d) Relative entropies between the steady-state distributions of the reduced and original models under different values of $\lambda$ when $r = 10$ (blue) and $r = 0$ (red).
(e) Simulations of the time-dependent mRNA distributions for the original model (blue) and the reduced model when $\lambda = 30$ (red) at different times.
(f) Relative entropies between the time-dependent mRNA distributions of the reduced model when $\lambda = 30$ and the original model as a function of time $t$ when the original model starts from the slow state $(0,0)$ (blue) and fast state $(1,0)$ (red). In (c)-(f), the model parameters are chosen as $a = 10, r = 10, v = 1, x = 0.8, y = 0.2$.}\label{transcription}
\end{center}
\end{figure}

We first consider the bursting kinetics of mRNAs. For simplicity, we assume that the gene switching rates $a_n = a$ and $b_n = b$ do not depends on the protein copy number $n$. If we focus on the dynamics of the mRNA, then the chemical state of the gene can be only described by two variables $(i,m)$: the gene activity $i$ and the mRNA copy number $m$. In this case, the standard model illustrated in Fig. \ref{model}(b) reduces to the Markov model illustrated in Fig. \ref{transcription}(a). To explain transcription bursts, we assume that $b\gg a$ and $s/b$ is finite. For convenience, let $\lambda = b/a\gg 1$ denote the ratio of the two gene switching rates. Then the leaving rates of all the chemical states are given by
\begin{equation*}
\begin{split}
q_{(0,m)} &= a+r+mv,\\
q_{(1,m)} &= b+s+mv = \lambda a(1+s/b)+mv,
\end{split}
\end{equation*}
which shows that
\begin{equation*}
\lim_{\lambda\rightarrow\infty}q_{(0,m)} < \infty,\;\;\;\lim_{\lambda\rightarrow\infty}q_{(1,m)} = \infty.
\end{equation*}
Thus all the inactive states $(0,m)$ are slow states and all the active states $(1,m)$ are fast states. The slow and fast state spaces are given by
\begin{equation*}
\begin{split}
A &= \set{(0,m):m=0,1,\cdots},\\
B &= \set{(1,m):m=0,1,\cdots},
\end{split}
\end{equation*}
and the generator matrix can be represented as the following block matrix:
\begin{equation*}
\Q = \left(\begin{array}{cccc;{2pt/2pt}cccc}
-q_{(0,0)} & r          &            &        & a          &            &            &        \\
v          & -q_{(0,1)} & r          &        &            & a          &            &        \\
           & 2v         & -q_{(0,2)} & \ddots &            &            & a          &        \\
           &            & \ddots     & \ddots &            &            &            & \ddots \\\hdashline[2pt/2pt]
b          &            &            &        & -q_{(1,0)} & s          &            &        \\
           & b          &            &        & v          & -q_{(1,1)} & s          &        \\
           &            & b          &        &            & 2v         & -q_{(1,2)} & \ddots \\
           &            &            & \ddots &            &            & \ddots     & \ddots
\end{array}\right).
\end{equation*}
According to our theory, the original model can be simplified to a reduced one by removal of the fast states. For simplicity, the state $(0,m)$ of the reduced model will be denoted by $m$.

More difficult is to calculate the effective transition rates of the reduced model. Let $\tilde{q}_{m,m'}$ denote the effective transition rate from state $m$ to state $m'$. If $m'<m$, it is easy to see that any fast transition path from $m$ to $m'$ much includes the transition $(1,m)\rightarrow (1,m-1)$. When $\lambda\gg 1$, the transition probability from state $(1,m)$ to state $(1,m-1)$ is given by
\begin{equation*}
w_{(1,m),(1,m-1)} = \lim_{\lambda\rightarrow\infty}\frac{mv}{\lambda a(1+s/b)+mv} = 0.
\end{equation*}
Thus all the fast transition paths from $m$ to $m'$ must have zero probability weights and have no contribution to the effective transition rate. As a result, the effective transition rate from $m$ to $m'$ is exactly the direct transition rate:
\begin{equation*}
\begin{split}
\tilde{q}_{m,m'} = q_{(0,m),(0,m')} =
\begin{cases}
mv,\;\;\;&m' = m-1,\\
0,\;\;\;&m' < m-1.
\end{cases}\end{split}
\end{equation*}
If $m'>m$, the only fast transition path $c$ from state $m$ to state $m'$ with positive probability weight is given by
\begin{equation*}
(0,m)\rightarrow(1,m)\rightarrow(1,m+1)\rightarrow\cdots\rightarrow(1,m')\rightarrow(0,m').
\end{equation*}
Intuitively, once is the gene becomes active, it can either synthesize an mRNA with probability $x = s/(b+s)$ or switch to the inactive state with probability $y = b/(b+s)$. When $\lambda\gg 1$, the transition probabilities along the path $c$ are given by
\begin{equation*}
\begin{split}
&w_{(0,m),(1,m)} = \frac{a}{a+r+mv},\\
&w_{(1,m),(1,m+1)} = \lim_{\lambda\rightarrow\infty}\frac{\lambda as/b}{\lambda a(1+s/b)+mv} = x,\\
&w_{(1,m'),(0,m')} = \lim_{\lambda\rightarrow\infty}\frac{\lambda a}{\lambda a(1+s/b)+m'v} = y,
\end{split}
\end{equation*}
and thus the probability weight of the path $c$ is
\begin{equation*}
\begin{split}
w_c &= w_{(0,m),(1,m)}w_{(1,m),(1,m+1)}\cdots w_{(1,m'),(0,m')}\\
&= \frac{ax^{m'-m}y}{a+r+mv}.
\end{split}
\end{equation*}
According to our theory, the effective transition rate from $m$ to $m'$ is the sum of the direct transition rate and the indirect transition rate via the fast transition path $c$:
\begin{equation*}
\begin{split}
\tilde{q}_{m,m'} &= q_{(0,m),(0,m')}+q_{(0,m)}w_c\\
&= \begin{cases}
r+axy,\;\;\;\;\;m' = m+1,\\
ax^{m'-m}y,\;\;\;m' > m+1.
\end{cases}\end{split}
\end{equation*}
This formula is informative. It indicates that the mRNA copy number may yield large jumps within a very short period, which exactly corresponds to random transcription bursts. Recall that $r$ is the basal transcription rate when the gene is inactive. If $r = 0$, it is easy to see that the burst size $m'-m$ follows the geometric distribution. A large basal transcription rate $r$ will lead to a deviation of the burst size from being geometrically distributed.

So far, we have calculated all the effective transition rates of the reduced model, as illustrated in Fig. \ref{transcription}(b). The reduced model includes long-range interactions of the mRNA copy number and naturally describes random transcription bursts. The bursts exactly correspond to fast transition paths of the original model, which are marked as red arrows in Fig. \ref{transcription}(a). Furthermore, we give the precise mathematical conditions for transcription bursts: $b\gg a$ and $s/b$ is finite. As pointed out intuitively by Paulsson \cite{paulsson2005models}: ``If genes are mostly inactive but transcribe a large number of mRNAs while in the active state, transcription could occur in bursts". Since the steady-state probability for a gene to be inactive is $P_{\textrm{inactive}} = b/(a+b) \approx 1$ and the transcription rate $s = \lambda a(s/b)$ in the active state is relatively large, our mathematical conditions are in full agreement with the above intuitive descriptions.

The occurrence of transcription bursts can be explained as follows. Since $b\gg a$, the time that the gene is active is much shorter compared to the time that the gene is inactive. Since $s/b$ is finite, when the gene is active, the large transcription rate $s$ will gives rise to the fast accumulation of the mRNA. Once the gene becomes inactive, the process of mRNA synthesis is terminated and mRNAs will be degraded until the gene becomes active again. Moreover, the mathematical conditions proposed here are consistent with recent single-cell experiments on transcription bursts of mammalian cells \cite{suter2011mammalian, bintu2016dynamics}. In \cite{suter2011mammalian}, the authors monitored the transcription kinetics in mouse fibroblasts by using single-cell time-lapse bioluminescence imaging. They found that the three parameters $a$, $b$, and $s$ for different genes are typically on the order of 0.01/min, 0.1/min, and 1/min, respectively (see Fig. 1D,E and Fig. S8 of \cite{suter2011mammalian} for details). These measurements agree reasonably well with our theory.

\subsection{Validity of the reduced mRNA model}
To validate the effectiveness of our simplification method, we numerically simulate both the original and reduced models using the Gillespie algorithm under a set of biologically relevant parameters. We first compare the steady-state behavior of the two models. Fig. \ref{transcription}(c) illustrates the steady-state distributions of the mRNA copy number for the two models. It can be seen that they agree with each other perfectly when $\lambda\gg 1$, but they fail as expected for smaller $\lambda$. In statistical physics and probability theory, the relative entropy (Kullback-Leibler divergence) is widely used to characterize the similarity of two probability distributions. The relative entropy between two probability distributions is zero if and only if they are exactly the same. Fig. \ref{transcription}(d) depicts the relative entropy between the steady-state distributions of the reduced and original models. It can be seen that the relative entropy decays dramatically when $\lambda$ is small and becomes close to zero when $\lambda\gg 1$. Recall that a large basal transcription rate $r$ will lead to a deviation of the burst size from the geometric distribution. To see the effect of the basal transcription rate on model simplification, we depict the relative entropies for $r = 0$ and $r = 10$ in Fig. \ref{transcription}(d), from which we see that a smaller $r$ will lead to a better approximation in the small $\lambda$ regime, but the approximation effect becomes almost the same when $\lambda\gg 1$.

To compare the dynamic behavior of the two models, we illustrate their time-dependent mRNA distributions in Fig. \ref{transcription}(e) when the two models start from the same slow state. When $\lambda\gg 1$, the two models exhibit almost the same dynamic behavior over the whole time axis. The situation is different when the original chain starts from a fast state. Let $\pai = (\pi_{(i,m)})$ denote the initial distribution of the original chain. When $m'\leq m$, the only fast transition path $c$ from state $(1,m')$ to state $(0,m)$ with positive probability weight is given by
\begin{equation*}
(1,m')\rightarrow(1,m'+1)\rightarrow\cdots\rightarrow(1,m)\rightarrow(0,m),
\end{equation*}
whose probability weight is $w_c = x^{m-m'}y$. According to our theory, the effective initial probability of state $m$ is the sum of the original initial probability and the contribution of all the fast transition paths:
\begin{equation*}
\begin{split}
\tilde\pi_m
&= \pi_{(0,m)}+\sum_{m'\leq m}\pi_{(1,m')}w_c \\
&= \pi_{(0,m)}+\sum_{m'\leq m}\pi_{(1,m')}x^{m-m'}y.
\end{split}
\end{equation*}
In particular, if the original chain starts from the fast state $(1,0)$, then the effective initial distribution is exactly the geometric distribution $\tilde\pi_m = x^my$.

To further clarify the effect of the initial distribution on model simplification, we depict the relative entropies between the time-dependent distributions of the reduced and original models in Fig. \ref{transcription}(f). The blue curve illustrates the case when the original model starts from the slow state $(0,0)$, from which we see that the probability distributions of the two models agree with each other perfectly over the whole time axis. In contrast, the red curve illustrates the case when the original model starts from the fast state $(1,0)$, from which we see that although the approximation causes large errors on a very short time scale, the probability distributions of the two models coincide with each other perfectly after time $t = 0.01$. These simulation results are in full agreement with our theory.

\subsection{Emergent translation bursts}
We next consider the bursting kinetics of proteins. To this end, we consider the full Markov model illustrated in Fig. \ref{model}(b). In single-cell experiments, it was consistently observed that the mRNA decays substantially faster relative to its protein counterpart \cite{shahrezaei2008analytical}. To explain translation bursts, we assume that $v\gg d$ and $u/v$ is finite. Let $\lambda = v/d\gg 1$ denote the ratio of the degradation rates of the mRNA and protein. Then the leaving rates of all the chemical states are given by
\begin{equation*}
\begin{split}
q_{(0,m,n)} &= a_n+r+mv+mu+nd \\
&= a_n+r+\lambda md(1+u/v)+nd,\\
q_{(1,m,n)} &= b_n+s+mv+mu+nd \\
&= b_n+s+\lambda md(1+u/v)+nd.
\end{split}
\end{equation*}
When $m\geq 1$, it is easy to see that
\begin{equation*}
\lim_{\lambda\rightarrow\infty}q_{(i,m,n)} = \infty.
\end{equation*}
This indicates that all the states $(i,m,n)$ with $m\geq 1$ are fast states and all the states $(i,0,n)$ with $m = 0$ are slow states. According to our theory, the original model can be simplified to a reduced one by removal of the fast states. For simplicity, the state $(i,0,n)$ of the reduced model will be denoted by $(i,n)$.
\begin{figure}[!htb]
\begin{center}
\centerline{\includegraphics[width=0.8\textwidth]{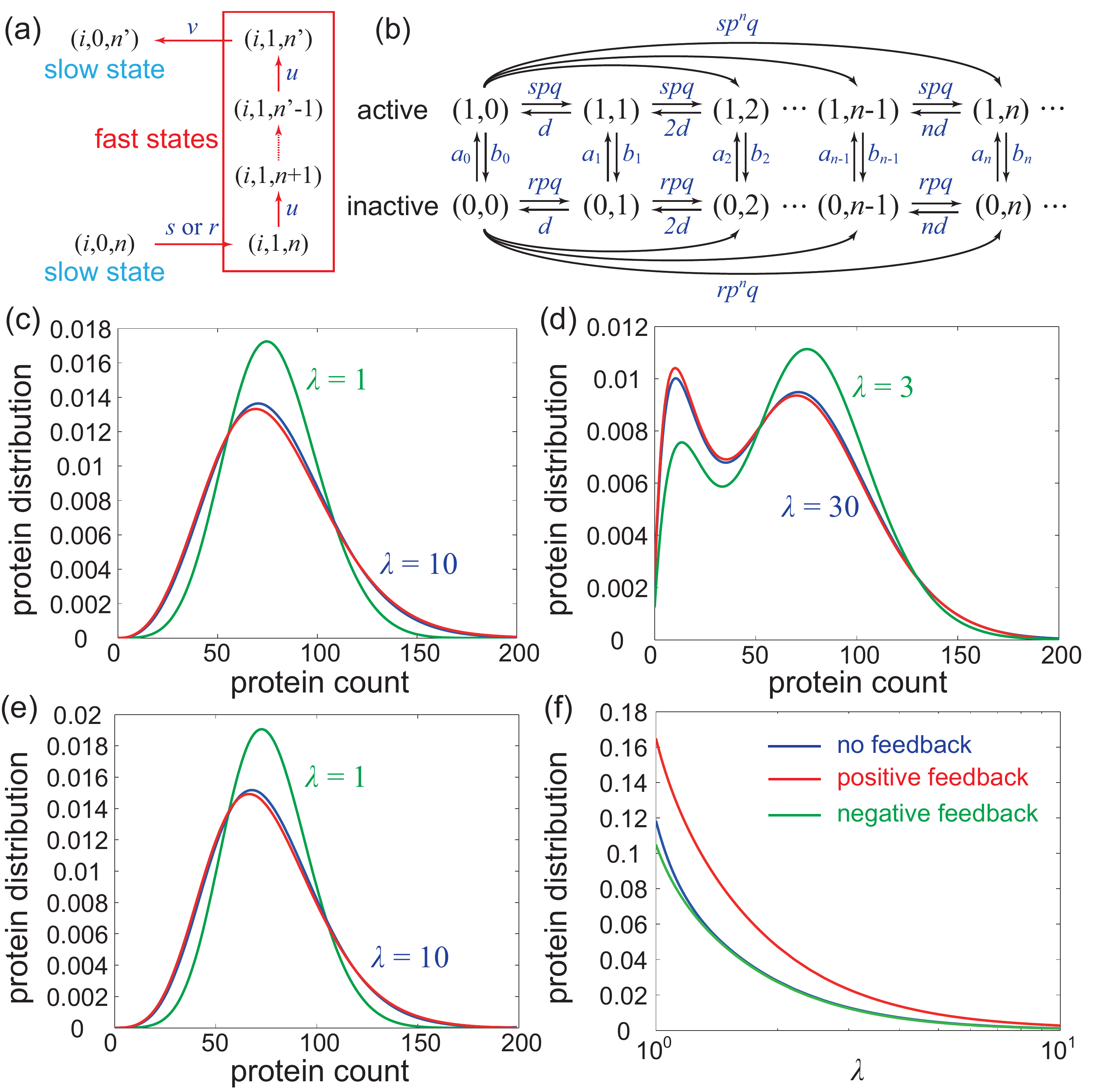}}
\caption{\textbf{Translation bursts as an emergent behavior from the fundamental biochemical reaction kinetics.}
(a) The only fast transition path from state $(i,n)$ to state $(i,n')$ with positive probability weight in the case of $n'>n$.
(b) The transition diagram of the reduced model when $v\gg d$ and $u/v$ is finite.
(c) Simulations of the steady-state protein distributions for the reduced model (red) and the original model when $\lambda = 1$ (green) and $\lambda = 10$ (blue) in networks with no feedback.
(d) Simulations of the steady-state protein distributions for the reduced model (red) and the original model when $\lambda = 3$ (green) and $\lambda = 30$ (blue) in networks with positive autoregulation.
(e) Simulations of the steady-state protein distributions for the reduced model (red) and the original model when $\lambda = 1$ (green) and $\lambda = 10$ (blue) in networks with negative autoregulation.
(f) Relative entropies between the steady-state protein distributions of the reduced and original models in three types of networks under different values of $\lambda$. In (c)-(f), the model parameters are chosen as $s = 10, r = 2, d = 1, p = 0.9, q = 0.1$. The gene switching rates are chosen as $a_n = 2.5, b_n = 0.5$ in the case of no feedback, $a_n = 0.05+n^2/2000, b = 0.5$ in the case of positive feedback, and $a = 2.5, b = 0.01+n^2/10000$ in the case of negative feedback.}\label{translation}
\end{center}
\end{figure}

We next calculate the effective transition rates of the reduced model. Let $\tilde{q}_{(i,n),(i',n')}$ denote the effective transition rate from state $(i,n)$ to state $(i',n')$. If $i'\neq i$ or $n'<n$, by using the criterion in Eq. \eqref{observation}, it is easy to check that all the fast transition paths from $(i,n)$ to $(i',n')$ have zero probability weight. This suggests that the effective transition rate from $(i,n)$ to $(i',n')$ is exactly the direct transition rate:
\begin{equation*}
\tilde{q}_{(i,n),(i',n')} = q_{(i,0,n),(i',0,n')}.
\end{equation*}
If $i'=i$ and $n'>n$, by using the criterion in Eq. \eqref{observation}, it is easy to check that the direct transition rate from $(i,n)$ to $(i,n')$ must be zero and the only fast transition path $c$ from $(i,n)$ to $(i,n')$ with positive probability weight is given by
\begin{equation*}
\begin{split}
(i,0,n)&\rightarrow(i,1,n)\rightarrow(i,1,n+1)\rightarrow\cdots\\
&\rightarrow(i,1,n')\rightarrow(i,0,n'),
\end{split}
\end{equation*}
as illustrated by the red arrow in Fig. \ref{translation}(a). Intuitively, once an mRNA is synthesized, it can either produce a protein with probability $p = u/(u+v)$ or be degraded with probability $q = v/(u+v)$. When $\lambda\gg 1$, the transition probabilities along the path $c$ are given by
\begin{equation*}
\begin{split}
&w_{(i,1,n),(i,1,n+1)} =
\lim_{\lambda\rightarrow\infty}\frac{\lambda mdu/v}{b+s+\lambda md(1+u/v)+nd} = p,\\
&w_{(i,1,n'),(i,0,n')} =
\lim_{\lambda\rightarrow\infty}\frac{\lambda md}{b+s+\lambda md(1+u/v)+n'd} = q,
\end{split}
\end{equation*}
Thus the effective transition rate from $(i,n)$ to $(i,n')$ is exactly the indirect transition rate via the fast transition path $c$:
\begin{equation*}
\begin{split}
\tilde{q}_{(i,n),(i,n')} &= q_{(i,0,n),(i,1,n)}w_{(i,1,n),(i,1,n+1)}\cdots w_{(i,1,n'),(i,0,n')}\\
&= \begin{cases}
rp^{n'-n}q,\;\;\;i = 0,\\
sp^{n'-n}q,\;\;\;i = 1.
\end{cases}\end{split}
\end{equation*}
This formula indicates that the protein copy number may yield large jumps within a very short period, which exactly corresponds to random translation bursts. In addition, the burst size $k = n'-n$ in both the active and inactive states follows the geometric distribution. The mean burst size is $\sum_{k=0}^\infty kp^kq = p/q$.

In \cite{cai2006stochastic}, the authors used single-molecular technologies to monitor real-time protein expression events in a single cell and found that the burst size is well fitted with the exponential distribution. In living cells, the mean burst size $p/q$ is relatively large, typically on the order of 100 for an \emph{E. coli} gene \cite{paulsson2005models}. In fact, the experimentally observed exponential distribution is exactly the continuous version of the theoretically derived geometric distribution. To see this, we assume that $p/q = bN$ with $N\gg 1$ being a scaling parameter. By taking the limit $N\rightarrow\infty$, while assuming $x = k/N$ to be a continuous variable, we have $p\rightarrow 1$ and $qN\rightarrow 1/b$. Thus the geometric distribution will converge to the exponential distribution:
\begin{equation*}
p^kq = qe^{k\log(1-q)} \approx qe^{-qk} \rightarrow \frac{1}{Nb}e^{-x/b}.
\end{equation*}

So far, we have calculated all the effective transition rates of the reduced model, as illustrated in Fig. \ref{translation}(b). The reduced model includes long-range interactions of the protein number and naturally describes random translation bursts. The random bursts exactly correspond to the fast transition paths of the original model, which are marked as red arrows in Fig. \ref{translation}(a). Furthermore, we give the precise mathematical conditions for translation bursts: $v\gg d$ and $u/v$ is finite. As pointed out intuitively by Paulsson \cite{paulsson2005models}: ``To truly have brief periods of high [protein] intensity it is not enough that $v\gg d$. To have true bursts in the model above, it is instead necessary that most cells have zero transcripts so that the total synthesis rate switches randomly from a low to a high value." According to our theory, the state space of the reduced model includes all the slow states $(i,0,n)$, which exactly corresponds to cells with zero transcripts. Therefore, our mathematical conditions are in full agreement again with the above intuitive descriptions.

The occurrence of translation bursts can be explained as follows. Since $v\gg d$ and $u/v$ is finite, the process of protein synthesis followed by mRNA degradation is essentially instantaneous. Once an mRNA is synthesized, the large translation rate $u$ will give rise to the fast accumulation of the protein. Once the mRNA is degraded, the process of protein synthesis is terminated and proteins will be degraded until another mRNA is synthesized again. Moreover, the mathematical conditions proposed here are consistent with recent bulk and single-cell experiments \cite{bernstein2002global, wang2002precision, hambraeus2003genome, bernstein2004global, lackner2007network, eden2011proteome, moran2012sizing, christiano2014global, schwalb2016tt}. Table \ref{halflife} lists recent measurements on the the medium mRNA half-life and medium protein half-life in different species, from prokaryotes to yeast then to higher eukaryotes. It can be seen that a typical protein half-life is about two orders of magnitude longer than an mRNA half-life in bacteria \cite{moran2012sizing} and is about one order of magnitude longer in yeast and mammals. These data agree reasonably well with our theory.
\begin{table}[!htb]
\centering
\begin{tabular}{|c|c|c|c|c|} \hline
Species           & Medium mRNA half-life & Reference & Medium protein half-life & Reference \\ \hline
\emph{Escherichia coli}  & 2 min - 6 min  & \cite{bernstein2002global, bernstein2004global}
& 20 hours        &\cite{moran2012sizing}      \\ \hline
\emph{Bacillus subtilis} & 4 min          &\cite{hambraeus2003genome}
& 20 hours        &\cite{moran2012sizing}      \\ \hline
Budding yeast     & 20 min         & \cite{wang2002precision}
& 8.8 hours       &\cite{christiano2014global} \\ \hline
Fission yeast     & 33 min         & \cite{lackner2007network}
& 12 hours        &\cite{christiano2014global} \\ \hline
Human             & 50 min         & \cite{schwalb2016tt}
& 9 hours         & \cite{eden2011proteome}    \\ \hline
\end{tabular}
\caption{Medium half-lifes or mRNAs and proteins in different species.}\label{halflife}
\end{table}

In addition, recent high-resolution fluorescence detection of tagged mRNAs in \emph{E. coli} cells determined that the number of transcripts per gene per cell ranges from 0.02 to 3 and the mean is only 0.4 (see Supplementary Table S6 in \cite{taniguchi2010quantifying} for details). This shows that most \emph{E. coli} cells have very low copies of transcripts. This is also well consistent with our theory.

\subsection{Validity of the reduced protein model}
In living cells, the products of many genes also act as their own transcription factors to form an autoregulatory gene network. This suggests that the gene switching rates $a_n$ and $b_n$ generally depend on the protein copy number $n$. There are three different situations that should be distinguished. If the network has no feedback, both $a_n$ and $b_n$ are constants independent of $n$. In a network with positive autoregulation, the protein binds to its own gene and actives its own transcription. In this case, $a_n$ is an increasing function of $n$ and $b_n$ is a constant. In contrast, in a network with negative autoregulation, the protein binds to its own gene and represses its own transcription. In this case, $a_n$ is a constant and $b_n$ is an increasing function of $n$.

To show the effectiveness of our simplification method, we numerically simulate both the original and reduced models using the Gillespie algorithm under a set of biologically relevant parameters. Fig. \ref{translation} (c)-(e) illustrate the steady-state distributions of the protein copy number for the two models in three types of gene networks. It can be seen that they coincide with each other reasonably well when $\lambda\gg 1$. However, the reduced model deviates from the original one in the small $\lambda$ regime. In networks with positive autoregulation, both the protein distributions of the two models may exhibit two different modes and give rise to bistability. However, the reduced model may reverse the heights of the two peaks when $\lambda$ is small, as can be seen from Fig. \ref{translation} (d). In fact, it was conjectured long ago by Thomas \cite{thomas1981relation} and was proved since then \cite{plahte1995feedback, gouze1998positive, snoussi1998necessary, cinquin2002positive, soule2003graphic} that the existence of a positive-feedback loop is a necessary condition for multistability. Our simulation results reinforce Thomas' conjecture again.

Fig. \ref{translation}(f) illustrates the relative entropy between the steady-state distributions of the reduced and original models in three types of networks. It can be seen that our approximation behaves completely different in the small $\lambda$ and large $\lambda$ regimes. Again, the relative entropy decays dramatically when $\lambda$ is small and becomes close to zero when $\lambda\gg 1$. Furthermore, our simplification approach shows similar approximation effects in networks with no feedback and negative feedback, but gives rise to a larger error in positive-feedback networks, which need a larger $\lambda$ to achieve the same approximation accuracy.

\section{Conclusions and discussion}\label{discussion}
In this paper, we develop an effective approach to simplify two-time-scale inhomogeneous continuous-time Markov chains with infinite state space by removal of the fast states. If the Markov chain has an infinite number of fast states, we are no longer able to remove the fast states one by one and calculate the inverse of an infinite-dimensional matrix. To overcome this difficulty, we introduce the key concept of fast transition paths and show that the effective transition rates (initial probabilities) of the reduced chain are the superposition of the direct transition rates (initial probabilities) and the indirect transition rates (initial probabilities) via all the fast transition paths.

There are two other innovative points of our simplification method. First, most of the previous papers focused on the homogeneous case when the generator matrix $\Q$ is independent of time $t$. Our strategy can be used to simplify inhomogeneous Markov chains whose generator matrices $\Q(t)$ are controlled by a time-dependent external parameter. Second, most of the previous papers considered singularly perturbed chains where the generator matrix $\Q$ depends linearly on the parameter $\lambda$ as $\Q = \lambda\tilde{\Q}+\hat{\Q}$, where $\tilde{\Q}$ and $\hat{\Q}$ are two generator matrices governing the fast and slow processes, respectively. However, in many areas of natural sciences such as biochemistry and biophysics, we frequently encounter Markov chains whose generator matrices do not depend linearly on $\lambda$. To see this, consider the scenario when the conformation of a receptor (DNA) is regulated by the binding of a ligand (transcription factor). In living cells, the ligand binding process in general cannot be viewed as an elementary reaction and thus the binding rate is usually modeled as a Hill function of the ligand concentration $\lambda$:
\begin{equation*}
q_{ij} = \frac{a\lambda^n}{\lambda^n+K^n},
\end{equation*}
with $n$ being the Hill coefficient. Even if the binding process can be viewed as an elementary reaction, if multiple ligands form a complex and bind to of the receptor, then the binding rate should be written as $q_{ij} = a\lambda^n$ with $n$ being the number of ligands in the complex. When the ligand concentration $\lambda\gg 1$, the traditional theory of singularly perturbed chains will become invalid but our theory is still applicable.

Another conceptual contribution of our simplification method is to clarify the probability implication of the effective initial distribution as the first passage distribution of the slow state space $A$. If the original chain starts from a slow state, the reduced chain should of course start from the same state and the probability distributions of the two chains will agree with each other over the whole time axis. However, if the original chain starts from a fast state, we show that although model simplification may cause large errors on a very short time scale, the probability distributions of the two chains will coincide with each other perfectly afterwards.

To validate the effectiveness of our theory, we apply it to the standard model of single-cell stochastic gene expression, which is a huge and infinite Markov chain on the three-dimensional lattice $\{(i,m,n):i = 0,1,m,n\geq 0\}$. It turns out that the concept of fast transition paths has important physical implications: these paths exactly correspond to random bursts during mRNA and protein synthesis. With this correspondence, we establish the rigorous mathematical theory of random gene expression bursts as an emergent behavior from the fundamental multi-scale biochemical reaction kinetics of stochastic gene expression. Furthermore, we give the mathematical conditions for both the transcriptional and translation bursts, which precisely coincide with recent bulk and single-cell experiments.

Of course, the applications of our simplification approach are not limited to the stochastic models of intracellular gene expression. We hope that the approach can be applied to more problems arising from physics, chemistry, and biology.

\section*{Acknowledgements}
The author gratefully acknowledges H. Qian, M.Q. Zhang, and M. Chen for stimulating discussions. The author is also grateful to the anonymous reviewers for their valuable comments and suggestions which help me greatly in improving the quality of this paper.

\setlength{\bibsep}{5pt}
\small\bibliographystyle{nature}
\bibliography{simplification}
\end{document}